\documentclass{PoS}
 \usepackage{bbold}
 
\usepackage{amssymb,amsmath,amsfonts,mathtools}
\usepackage[sort,compress]{cite}
\usepackage[bulletsep]{collref}
\usepackage{caption}
\usepackage{subcaption}
\usepackage{booktabs}
\usepackage{dsfont}

\usepackage{color}

\def \adss {$AdS_5 \times S^5$\ }

\newcommand{\be}{\begin{equation}}
\newcommand{\ee}{\end{equation}}

\newcommand{\beq}{\begin{equation}}
\newcommand{\eeq}{\end{equation}}

\renewcommand{\S}{\Sigma}

\newcommand{\Tr}{\textup{Tr}}

\title{
\vspace*{-1.5cm}
{\normalsize {\rm \hfill{HU-EP-17/06} }} \\
\vspace*{1.5cm}
Strings on the lattice and AdS/CFT}
 
\ShortTitle{Strings on the lattice and AdS/CFT}

\author{\speaker{Valentina Forini} \\ %\thanks{A footnote may follow.}\\
        Humboldt-Universit\"at zu Berlin, Zum Gro\ss en Windkanal 6, 12489 Berlin, Germany\\
        E-mail: \email{forini@physik.hu-berlin.de}}

\author{Lorenzo Bianchi\\
       II. Institut f\"ur Theoretische Physik, Universit\"at Hamburg, Luruper Chaussee 149, 22761 Hamburg, Germany  \\
        E-mail: \email{lorenzo.bianchi@desy.de}}

\author{Bjoern Leder\\
        Institutsrechenzentrum, Institut f\"ur Physik\\ 
        Humboldt-Universit\"at zu Berlin, Newtonstr. 15, 12489 Berlin, German\\
        E-mail: \email{leder@physik.hu-berlin.de}}
        
\author{Philipp T\"opfer\\
          Humboldt-Universit\"at zu Berlin, Zum Gro\ss en Windkanal 6, 12489 Berlin, Germany\\
        E-mail: \email{philipp.toepfer@physik.hu-berlin.de}}

\author{Edoardo Vescovi\\
          Humboldt-Universit\"at zu Berlin, Zum Gro\ss en Windkanal 6, 12489 Berlin, Germany\\
          Institute of Physics, University of S\~{a}o Paulo, Rua do Mat\~{a}o 1371, 05508-090 S\~{a}o Paulo, Brazil\\
        E-mail: \email{vescovi@if.usp.br}}                

\abstract{We present a new auxiliary field representation for the four-fermi term of the gauge-fixed Green-Schwarz superstring action which describes fluctuations around the null-cusp background in~$AdS_5\times S^5$. 
We sketch the main features of the fermionic operator spectrum, identifying the region of parameter space where the sign ambiguity is absent.
Measurements for the observables in the setup here described are presented and discussed in~\cite{toappear}.}

\FullConference{34th annual International Symposium on Lattice Field Theory\\ 24-30 July 2016\\
University of Southampton, United Kingdom
}

\begin{document}

\section{Discussion}
\label{sec:discussion}

Investigating via lattice field theory methods quantum string worldsheets relevant in AdS/CFT has recently become a concrete possibility~\cite{Roiban, Forini:2016sot,Bianchi:2016cyv}~\footnote{For lattice investigations in other models relevant in AdS/CFT see e.g.~\cite{Catterall_physrept, Schaich:2015ppr,Bergner:2016sbv, Schaich:2016jus, Berkowitz:2016jlq}. }.  The model under study is the AdS-lightcone gauge-fixed (Type IIB Green-Schwarz)  superstring action~\cite{MTMTT} describing fluctuations around the classical string solution ending, at the AdS boundary, on a lightlike cusped Wilson loop~\cite{Giombi}. It consists of a highly non-trivial two-dimensional quantum field theory with quartic fermionic interactions; its perturbative analysis provides, according to AdS/CFT, strong coupling information on the behaviour of several important observables in the dual gauge theory %~\cite{Alday:2010ku, Basso:2013vsa,Basso:2013aha,Fioravanti:2015dma, 
%Bonini:2015lfr} 
and has been explicitly performed up to two loops  in sigma-model semiclassical quantization~\cite{Giombi}. 

In~\cite{Roiban, Forini:2016sot,Bianchi:2016cyv} lattice simulations employing a Rational Hybrid Monte Carlo (RHMC) algorithm were performed  in order to measure the vacuum expectation value of the action -- from which the cusp anomaly of $\mathcal{N}=4$ super Yang-Mills~\cite{Roiban, Forini:2016sot,Bianchi:2016cyv} can be extracted -- and the mass~\cite{Forini:2016sot,Bianchi:2016cyv} of the two AdS excitations transverse to the relevant null cusp classical string solution. For both these observables it is possible to compare the results with the  behavior predicted via integrability~\cite{BES,Basso} at various values of the coupling~$g= \sqrt{\lambda}/(4\pi)$. At large $g$, \emph{i.e.} the perturbative regime of the sigma-model,   for both the observables a good agreement was found. At lower values of $g$, after a non-perturbative subtraction of quadratic divergences, a \emph{complex} phase in the fermion determinant -- in fact, a Pfaffian -- was detected~\cite{Forini:2016sot,Bianchi:2016cyv}.  There, it was also concluded that a strong sign problem in simulations would appear for values of the (lattice) coupling $g\leq5$.

The origin of such complex phase is a non-hermitian piece in the Lagrangian, a specific Yukawa-like term resulting from the standard linearization of a quartic fermionic interaction which appears originally as a ``repulsive'' potential~\cite{Bianchi:2016cyv}.  
In this contribution we present a new auxiliary field representation of the four-fermi term, following  an algebraic manipulation of the original fermionic Lagrangian inspired by~\cite{Catterall:2015zua} (see also~\cite{Catterall:2016dzf}). The result is a Lagrangian linear in fermions which is fully hermitian, and a quadratic fermionic operator $O_F$  which  -- in presence or not of Wilson-like terms -- is antisymmetric and obeys a constraint reminiscent of the ``$\gamma_5$-hermiticity'' in lattice QCD.  These two properties ensure that $\det O_F$ is real and non-negative, from which a \emph{real} Pfaffian $({\rm Pf}\,O_F)^2=\det O_F \geq0$. Eliminating the complex phase allows us to eliminate a systematic error in measurements~\footnote{In order to treat it via standard reweighting, the phase should be calculated explicitly. As it is highly non trivial to evaluate efficiently complex determinants for arbitrarily big matrices, in~\cite{Bianchi:2016cyv} this was done only for small lattices, \emph{i.e.} small values of lattice points $N$.  Observing  that the reweighting had no effect on the central value of the observables  under study, the phase was omitted from the simulations in order to  consistently take the continuum limit ($N\to\infty$), and in absence of data for  larger lattices the possible systematic error related to this procedure was not assessed.}. As a sign ambiguity remain in its definition, ${\rm Pf} \,O_F=\pm\sqrt{\det O_F}$, the non definite positive Pfaffian can not be directly treated using a (rational) hybrid Monte Carlo algorithm, but only upon replacement ${\rm Pf} \,O_F \rightarrow \det (O_F^\dagger O_F)^{\frac{1}{4}}$ and a reweighting procedure which can potentially break down if the sign ambiguity is severe. Below, after presenting the details of the new linearization, we show numerically some relevant features of the spectrum of the new quadratic fermionic operator. The origin of the sign ambiguity appears to be related to the Yukawa-like terms, including those present in the original Lagrangian (before linearization). Also, one may identify the region of parameter space where no sign flips can occur, and that therefore cannot be affected by a sign problem. As the latter occurs at $g\sim 2$ and lower, this leaves the possibility of ``safe'' measurements at e.g. $g=10$, where features of the non-perturbative regime appear to be detected by (previous~\cite{Bianchi:2016cyv} and) current~\cite{toappear} simulations. The measurements of relevant observables in this novel setup are discussed in the forthcoming~\cite{toappear}.

\section{Linearization}
\label{sec:observable}
The euclidean superstring action  in AdS-lightcone gauge-fixing~\cite{MTMTT}  describing quantum  fluctuations around  the null-cusp background  in \adss reads~\cite{Giombi}
\begin{eqnarray}\nonumber
&& \!\!\!\!\!\!\!\!\!\!\!\!
S_{\rm cusp}=g \int dt ds~ \Big\{ 
%\mathcal{L}_{\rm cusp}\\\nonumber
%&& \!\!\!\!\!\! 
%\mathcal{L}_{\rm cusp} = 
|\partial_{t}x+\textstyle{\frac{1}{2}}x|^{2}+\frac{1}{ {z}^{4}} |\partial_{s} {x}-\textstyle{\frac{1}{2}} {x}|^{2}+\left(\partial_{t}z^{M}+\frac{1}{2} {z}^{M}+\frac{i}{ {z}^{2}} {z}_{N} {\eta}_{i}\left(\rho^{MN}\right)_{\phantom{i}j}^{i} {\eta}^{j}\right)^{2}\\\label{S_cusp}
 && \!\!\!\!\!\!\!\!\!\!\!\!
+\frac{1}{ {z}^{4}}\left(\partial_{s} {z}^{M}-\textstyle{\frac{1}{2}} {z}^{M}\right)^{2}  %\\\label{S_cusp}
  +i\left( {\theta}^{i}\partial_{t}{\theta}_{i}+ {\eta}^{i}\partial_{t}{\eta}_{i}+ {\theta}_{i}\partial_{t}{\theta}^{i}+ {\eta}_{i}\partial_{t} {\eta}^{i}\right)-\textstyle{\frac{1}{{z}^{2}}}\left( {\eta}^{i}{\eta}_{i}\right)^{2}  \\\nonumber
 &&  \!\!\!\!\!\!\!\!\!\!\!\!
 +2i\Big[\textstyle{\frac{1}{z^{3}}}z^{M} {\eta}^{i}\left(\rho^{M}\right)_{ij}
 \left(\partial_{s} \theta^j-\textstyle{\frac{1}{2}} \theta^j-\frac{i}{{z}} {\eta}^{j}\left(\partial_{s} {x}-\frac{1}{2} {x}\right)\right)
 %\\
% &&  \!\!\!\!\!\!
\textstyle{+\frac{1}{{z}^{3}}{z}^{M}{\eta}_{i} (\rho_{M}^{\dagger} )^{ij}\left(\partial_{s}{\theta}_{j}-\frac{1}{2}{\theta}_{j}+\frac{i}{{z}}{\eta}_{j}\left(\partial_{s}{x}-\frac{1}{2}{x}\right)^{*}\right)\Big]\,\Big\}}\,
\end{eqnarray}
where $x,x^*$ are two bosonic fields transverse  to the subspace $AdS_3$  of the classical solution and $z^M\, (M=1,\cdots, 6)$ are the bosonic fileds of the $AdS_5\times S^5$ spacetime in Poincar\'e patch, with  $z=\sqrt{z_M z^M}$.  %As mentioned above,  the lagrangean above neither contains gauge fields nor actual fermions. Indeed, 
 The Gra\ss mann-odd fields $\theta_i,\eta_i,\, i=1,2,3,4$ are complex variables (no Lorentz spinor indices appear), such that $\theta^i = (\theta_i)^\dagger,$ $\eta^i = (\eta_i)^\dagger$ and  transforming in the fundamental representation of the $SU(4)$ R-symmetry group. The matrices $\rho^{M}_{ij} $ are the off-diagonal
blocks of $SO(6)$ Dirac matrices $\gamma^M$ in  chiral representation, 
%~\footnote{$(\rho_{M}^{\dagger})^{ij}$ is here indicating the upper off-diagonal block $(\rho^M)^{ij}$ (carrying \emph{upper} indices). The six $4\times 4$ matrices $(\rho^M)_{ij}$ represent the off-diagonal blocks of the SO(6), $8\times 8$ Dirac matrices $\gamma^M$ in the chiral representation
%\begin{equation} 
%\gamma^M\equiv \begin{pmatrix}
%0  & \rho^\dagger_M   \\
% \rho^M   &  0 
%\end{pmatrix}
%=
%\begin{pmatrix}
%0  & (\rho^M)^{ij}   \\
%(\rho^M)_{ij}   &  0 
%\end{pmatrix}
%\end{equation}
%The two off-diagonal blocks, carrying upper and lower indices respectively, are related by $(\rho^M)^{ij}=-(\rho^M_{ij})^*\equiv(\rho^M_{ji})^*$, so that indeed the block with upper indices is the conjugate transpose of the block with lower indices.
%} 
and
$(\rho^{MN})_i^{\hphantom{i} j} = (\rho^{[M} \rho^{\dagger N]})_i^{\hphantom{i} j}$ are  the
$SO(6)$ generators. 
%The fields $z^M$ are neutral under U(1), $\theta^i$ and $\eta^i$ have opposite
%charges and the charge of $\eta_i$ is half the charge of $x$. 
In the action \eqref{S_cusp} a massive parameter ($\sim P_+$) is missing, which we restore below in \eqref{Scuspquadratic} defining it as~$m$. 
% and introducing its dimensionless counterpart $M=a\,m$, where $a$ is the lattice spacing.  
%We emphasize that, in \eqref{S_cusp},  local bosonic (diffeomorphism) and fermionic ($\kappa$-) symmetries originally present have been fixed.}
 
%
As standard, to take into account the fermionic contribution in the case of higher-order interactions one linearizes the corresponding Lagrangian  and then formally integrates out  the Gra\ss mann-odd fields letting their determinant - here, a Pfaffian -  to enter  the Boltzmann weight of each configuration through re-exponentiation
\begin{equation}\label{fermionsintegration}
 \int \!\! D\Psi~ e^{-\textstyle\int dt ds \,\Psi^T O_F \Psi}={\rm Pf}\,O_F\longrightarrow (\det O_F\,O^\dagger_F)^{\frac{1}{4}}= \int \!\!D\xi D\bar\xi\,e^{-\int dt ds\, \bar\xi(O_FO^\dagger_F)^{-\frac{1}{4}}\,\xi}~
 \end{equation}
 where the replacement is needed in the case of non positive definite Pfaffian. 
% the straightforward way to linearize the quartic fermionic interactions in~\eqref{S_cusp} is to introduce  a set of $7$ real auxiliary  fields, one scalar $\phi$ and a  $SO(6)$ vector field $\phi_M$ as in~\colb{ROIBAN,CITEUS}
%\begin{eqnarray}\label{HubbardStratonovich}
%&& \!\!\!\!\!\!\!
%\exp \Big\{-g\int dt ds  \Big[-\textstyle{\frac{1}{{z}^{2}}}\left( {\eta}^{i}{\eta}_{i}\right)^{2}  +\Big(\textstyle{\frac{i}{ {z}^{2}}} {z}_{N} {\eta}_{i}{\rho^{MN}}_{\phantom{i}j}^{i} {\eta}^{j}\Big)^{2}\Big]\}\\\nonumber
%&& 
%\sim\,\int D\phi D\phi^M\,\exp\Big\{-  g\int dt ds\,[\textstyle\frac{1}{2}{\phi}^2+\frac{\sqrt{2}}{z}\phi\,\eta^2 +\frac{1}{2}({\phi}_M)^2-i\,\frac{\sqrt{2}}{z^2}\phi^M {z}_{N} \,\big(i \,{\eta}_{i}{\rho^{MN}}_{\phantom{i}j}^{i} {\eta}^{j}\big)]\Big\}~.
%\end{eqnarray}
%The second quartic interaction above squares an hermitian bilinear, $\Big(i\,\eta_i {\rho^{MN}}^i{}_j \eta^j\Big)^\dagger=i\eta_j\,{\rho^{MN}}^j{}_i\,\eta^i$, and comes in the exponential as a ?repulsive? potential. This has the final effect of an imaginary part in the auxiliary
%
% due to $\Big(i\,\eta_i {\rho^{MN}}^i{}_j \eta^j\Big)^\dagger=i\eta_j\,{\rho^{MN}}^j{}_i\,\eta^i$, the corresponding Yukawa term makes the linearized Lagrangian not hermitian. A complex phase is then easily detected in the resulting Pfaffian , and the the  only  question being whether the latter is treatable via standard reweighing. Below we find evidence that %, at least in this setting, 
%this is not the case.
Here, we focus on the part of the Lagrangian in \eqref{S_cusp} which is quartic in fermion
 \begin{equation}\label{eq:quarticaction}
  \mathcal{L}_4=\frac{1}{z^2}\left[- (\eta^2)^2+\left(i\, \eta_i {(\rho^{MN})^i}_j n^N \eta^j\right)^2\right]
\end{equation}
where $n^M=\frac{z^M}{z}$. Notice the plus sign in front of the second term in \eqref{eq:quarticaction}, which squares an hermitian bilinear $(i\,\eta_i {\rho^{MN}}^i{}_j \eta^j)^\dagger=i\eta_j\,{\rho^{MN}}^j{}_i\,\eta^i$~\cite{Bianchi:2016cyv}. Then the  standard Hubbard-Stratonovich transformation
\begin{eqnarray}\label{HubbardStratonovich}
&& \!\!\!\!\!\!\!
\exp \Big\{-g\int dt ds  \Big[-\textstyle{\frac{1}{{z}^{2}}}\left( {\eta}^{i}{\eta}_{i}\right)^{2}  +\Big(\textstyle{\frac{i}{ {z}^{2}}} {z}_{N} {\eta}_{i}{\rho^{MN}}_{\phantom{i}j}^{i} {\eta}^{j}\Big)^{2}\Big]\}\\\nonumber
&& 
\sim\,\int D\phi D\phi^M\,\exp\Big\{-  g\int dt ds\,[\textstyle\frac{1}{2}{\phi}^2+\frac{\sqrt{2}}{z}\phi\,\eta^2 +\frac{1}{2}({\phi}_M)^2-i\,\frac{\sqrt{2}}{z^2}\phi^M {z}_{N} \,\big(i \,{\eta}_{i}{\rho^{MN}}_{\phantom{i}j}^{i} {\eta}^{j}\big)]\Big\}~
\end{eqnarray}
generates a non-hermitian term, the last one above, resulting in a complex-valued Pfaffian for the fermionic operator. Here we provide a solution to this problem by rewriting the Lagrangian \eqref{eq:quarticaction} in a suitable form. The procedure is inspired by~\cite{Catterall:2015zua}, where a simpler  action with SO(4) four-fermion terms  in three dimensions was considered (see also the four-dimensional $SU(4)$ counterpart in~\cite{Catterall:2016dzf}). The Lagrangian \eqref{eq:quarticaction} is invariant under $SU(4)\times U(1)$ transformations and this requires a generalization of \cite{Catterall:2015zua}. Let us start by eliminating the matrices $\rho^{MN}$ from the second term of \eqref{eq:quarticaction} in favour of $\rho^M$. After some $\rho$-matrices manipulations we get
\begin{align}
\mathcal{L}_4=\frac{1}{z^2}\left(- 4\, (\eta^2)^2+2\left|\eta_i (\rho^N)^{ik} n_N \eta_k\right|^2\right)
\end{align}
where the plus sign in front of the second term still prevents a real Pfaffian after the Hubbard-Stratonovich transformation. We then define a duality transformation, reminiscent of the standard Hodge duality, but adapted to our particular case. Given ${\S_i}^j\equiv\eta_i \eta^j$ the dual matrix $\tilde{\S}_j{}^i$ is defined by
\begin{align}
\tilde{\S}_j{}^i=n_N n_L(\rho^N)^{ik}(\rho^L)_{jl} {\S_k}^l
\end{align}
Notice that $\tilde{\tilde \S}=\S$ and ${\S^i}_j\equiv ({\S_i}^j)^\dagger={\S_j}^i $. One can then easily rewrite the quartic Lagrangian as
\begin{align}\label{lagsig}
 \mathcal{L}_4=\frac{2}{z^2}\Tr\left( \S\S+ \tilde \S\tilde \S- \S\tilde \S\right)
\end{align}
where the trace is over $SU(4)$ fundamental indices. Although we split the first two terms in \eqref{lagsig} to exhibit the neutrality of the Lagrangian under duality transformation it is useful to keep in mind that $\Tr \tilde \S\tilde \S=\Tr \S\S$. Since we want to write down a Lagrangian as the sum of two terms squared, it is natural to introduce the self- and antiself-dual part of $\S$ 
\begin{align}
 {\S_{\pm}}=\S \pm \tilde \S
\end{align}
such that $\tilde\S_{\pm}=\pm \S_{\pm}$. Now the crucial, though elementary fact that $\Tr \S_{\pm}\S_{\pm}=2\Tr \left(\S\S\pm  \S\tilde \S\right)$ gives us some freedom in the choice of the sign in the Lagrangian, since
\begin{align}\label{lagpm}
 \mathcal{L}_4=\frac{1}{z^2}\Tr\left(4 \S\S\mp \S_{\pm} \S_{\pm} \pm 2 \S\S \right)~.
\end{align}
This last equation proves that the sign problem was an artefact of our naive linearization. Indeed \eqref{lagpm} provides two equivalent forms of the same action, one which would lead to a sign problem and one which would not. Choosing the latter, i.e. the one involving $\S_+$, we get
 \begin{equation}
 \mathcal{L}_4=\frac{1}{z^2}\left(- 6\, (\eta^2)^2 - {\S_{+}}_i^j{\S_{+}}_j^i \right)~.
\end{equation}
In this form the Lagrangian is suitable for a Hubbard-Stratonovich transformation. In particular we have
\begin{multline}\label{HubbardStratonovich}
\exp \Big\{-g\int dt ds  \Big[-\textstyle{\frac{1}{z^2}\left(- 6\, (\eta^2)^2 - {\S_{+}}_i^j{\S_{+}}_j^i \right)}\Big]\Big\}\\\nonumber
\sim ~\int D\phi D\phi^M\,\exp\Big\{-  g\int dt ds\,[\textstyle \frac{12}{z} \eta^2 \phi +6\phi^2+\frac{2}{z} {\S_+}^i_j \phi^j_i +\phi^i_j \phi^j_i ]\Big\}~,
\end{multline}
where $\phi^i_j$ can be thought of as a $4\times 4$ complex hermitian matrix with 16 real degrees of freedom. Therefore the new linearization proposed here introduces a total of 17 auxiliary fields.

The final form of the Lagrangian is then
\begin{eqnarray}\label{Scuspquadratic}
{\cal L} &=&  {| \partial_t {x} + {\frac{m}{2}}{x} |}^2 + \frac{1}{{ z}^4}{\big| \partial_s {x} -\frac{m}{2}{x} |}^2
+ (\partial_t {z}^M + \frac{m}{2}{z}^M )^2 + \frac{1}{{ z}^4} (\partial_s {z}^M -\frac{m}{2}{z}^M)^2
\cr
&+&6\phi^2+\phi^i_j \phi^j_i+\psi^T O_F \psi\
\end{eqnarray}
with $\psi\equiv\left(\theta^{i},\theta_{i},\eta^{i},\eta_{i}\right)$ and
%%%%%%%%%%%%%%%%%%%%%%%%%%%%%%%%%%%%%%%%%%%%%
\begin{equation} \label{OF}
O_F =\left(\begin{array}{cccc}
0 & \mathrm{i}\partial_{t} & -\mathrm{i}\rho^{M}\left(\partial_{s}+\frac{m}{2}\right)\frac{{z}^{M}}{{z}^{3}} & 0\\
\mathrm{i}\partial_{t} & 0 & 0 & -\mathrm{i}\rho_{M}^{\dagger}\left(\partial_{s}+\frac{m}{2}\right)\frac{{z}^{M}}{{z}^{3}}\\
\mathrm{i}\frac{{z}^{M}}{{z}^{3}}\rho^{M}\left(\partial_{s}-\frac{m}{2}\right) & 0 & 2\frac{{z}^{M}}{{z}^{4}}\rho^{M}\left(\partial_{s}{x}-m\frac{{x}}{2}\right) & \mathrm{i}\partial_{t}-A^{T}\\
0 & \mathrm{i}\frac{{z}^{M}}{{z}^{3}}\rho_{M}^{\dagger}\left(\partial_{s}-\frac{m}{2}\right) &\mathrm{i}\partial_{t}+A & -2\frac{{z}^{M}}{{z}^{4}}\rho_{M}^{\dagger}\left(\partial_{s}{x}^\ast-m\frac{{x}}{2}^\ast\right)
\end{array}\right)~,
\end{equation}
where
\begin{align}
A&=-\frac{6}{z}\phi + \frac{1}{z}\tilde{\phi}+\frac{1}{z^{3}}\rho^\ast_{N}\tilde{\phi}^{T}\rho^{L}z^{N}z^{L}+\mathrm{i}\frac{z^{N}}{z^2}\rho^{MN}\partial_{t}z^{M}, \\
\tilde{\phi}&\equiv \left(\tilde{\phi}_{ij}\right)\equiv \left(\phi^{i}_{j}\right).
\end{align}
%

%In the simpler cases of Refs.~\cite{Catterall:2015zua,Catterall:2016dzf},  a suitable choice of  Yukawa terms turns out to ensure a definite positive Pfaffian, in connection with the relevant operator being real and antisymmetric and a double degeneracy observed in the spectrum, see comments in the next section.   
%%%%%%%%%%%%%%%%%%%%%%%%%%%%%%%%%%%%%%%%%%%%%

\section{The spectrum of the fermionic operator}
\label{sec:simulation}
 %
%As mentioned above, 
In simpler cases of models with four-fermion interactions~\cite{Catterall:2015zua,Catterall:2016dzf} a choice of  Yukawa terms  similar in spirit to the one described in the previous section turns out to ensure a definite positive Pfaffian. 
There the relevant operator is real and antisymmetric -- so that its purely imaginary eigenvalues come in pairs $(i\,a,-ia)$ -- and the key point there is that eigenvalues are also doubly degenerate. One may then define the Pfaffian as the product of eigenvalues with positive imaginary part on the initial configuration. As the simulation progresses, sign flips in the Pfaffian correspond to an odd number of eigenvalues crossing through the origin, but as all eigenvalues are doubly degenerate such sign changes cannot occur.

\begin{figure}[h]
   \centering
 \includegraphics[scale=0.7]{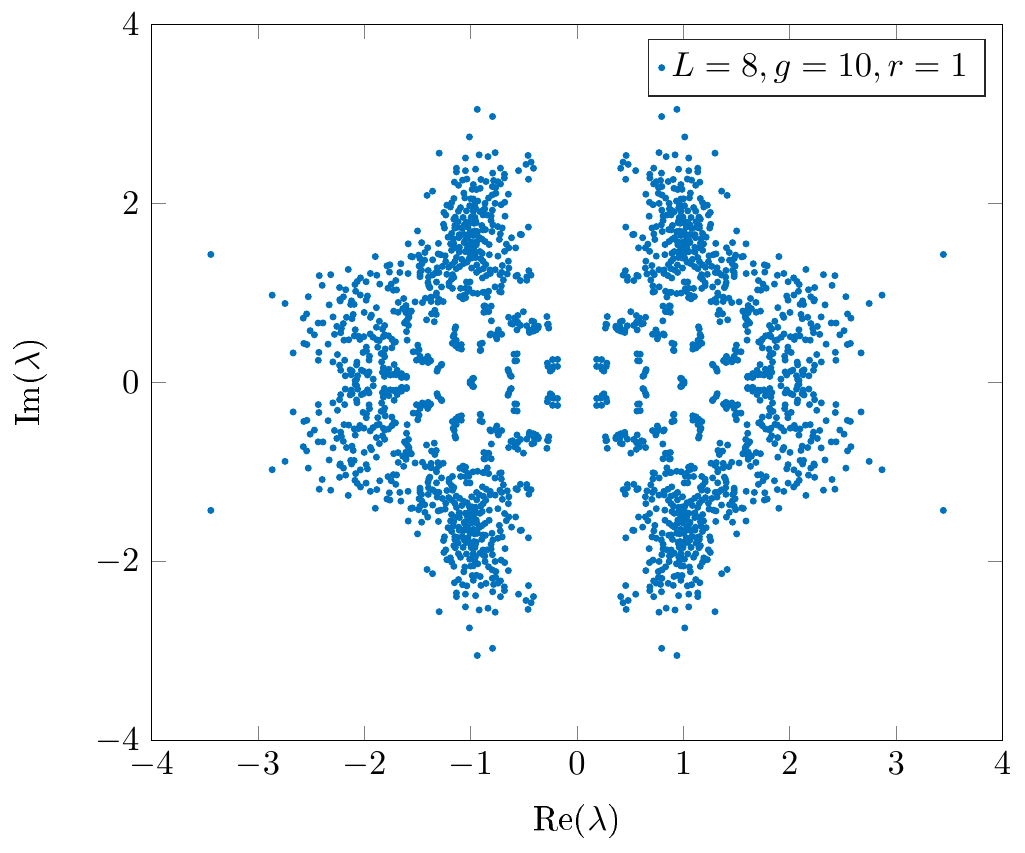}
 \includegraphics[scale=0.7]{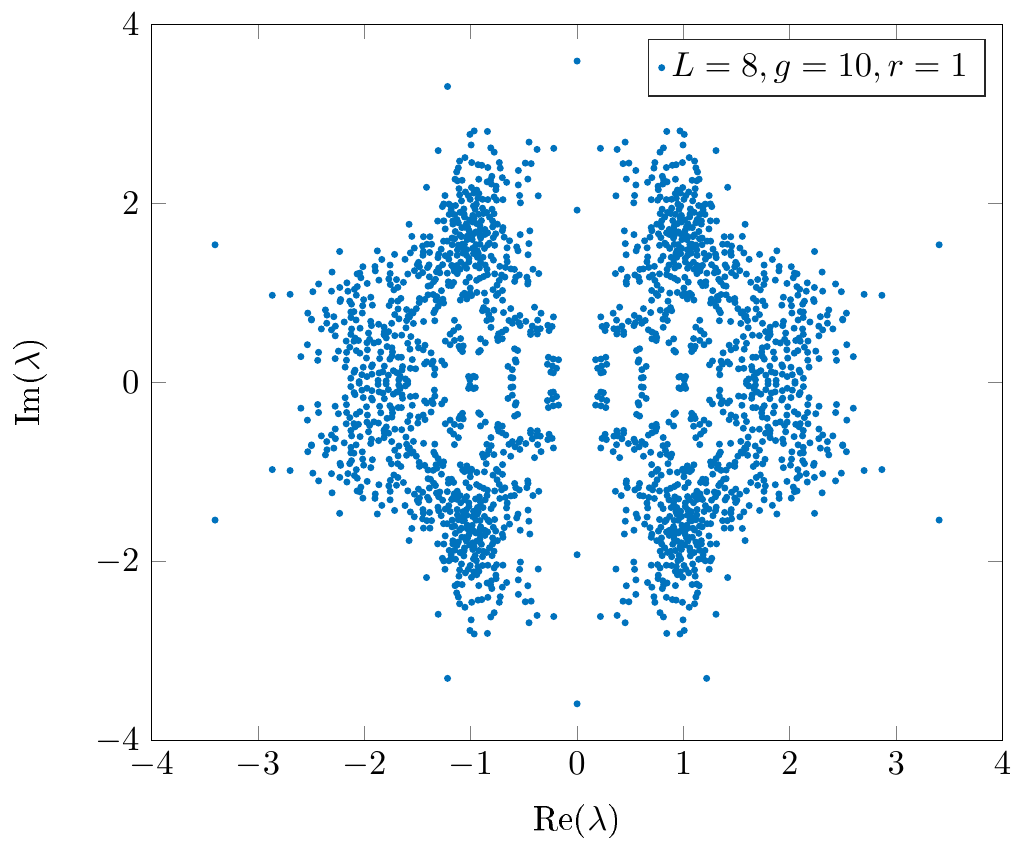}
 \includegraphics[scale=0.7]{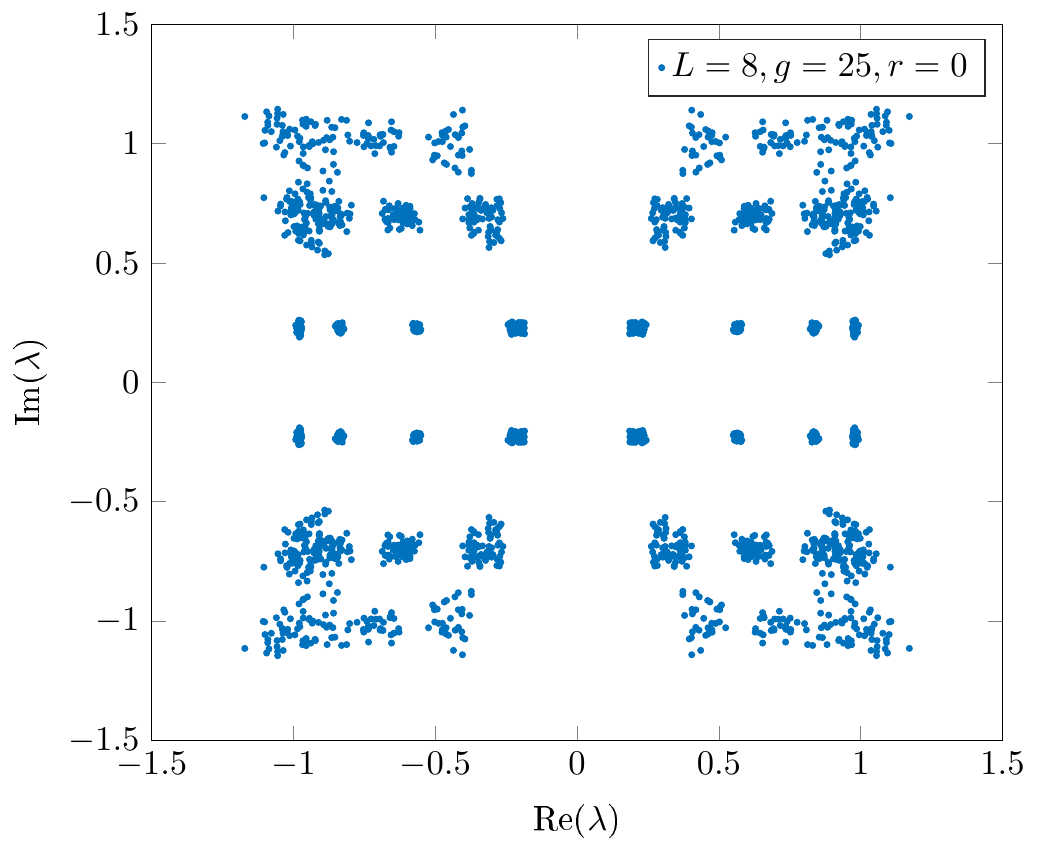}
 \includegraphics[scale=0.7]{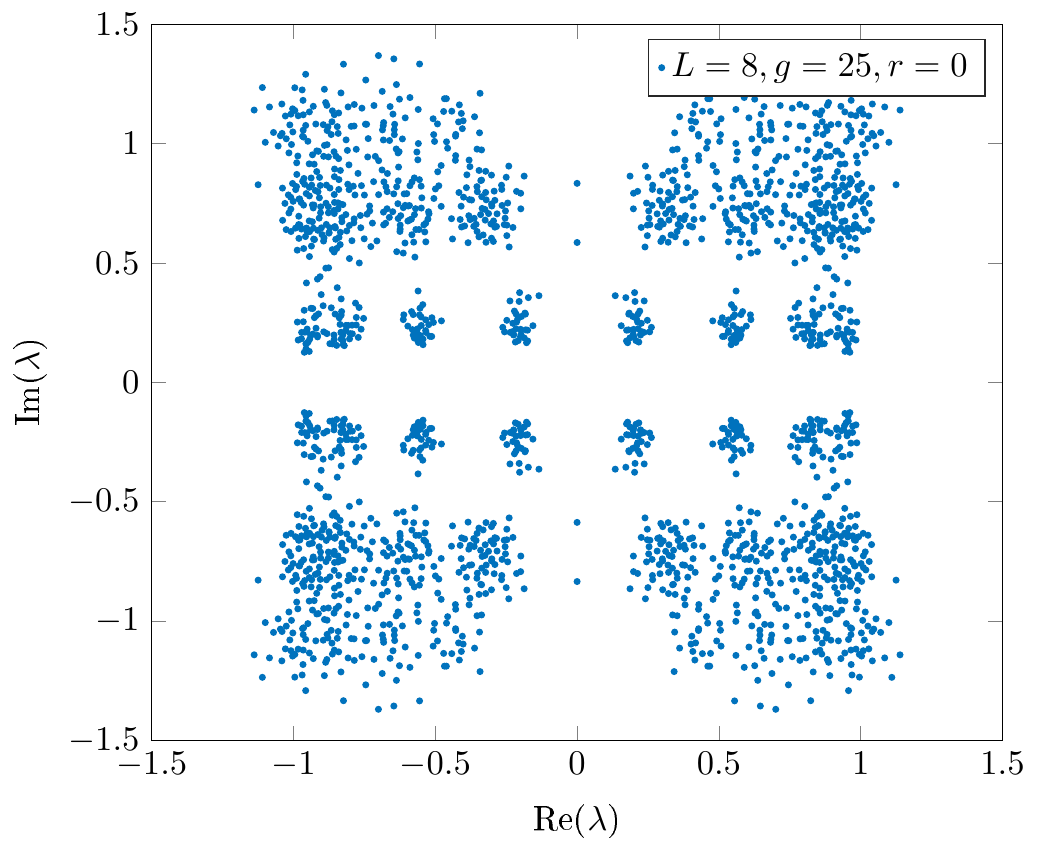}
 \caption{Spectrum of $O_F$, in absence (left diagrams) and presence (right diagrams) of  $A$ (Yukawa-like) terms.}
\label{fig:spectrum_ferm}
\end{figure}

Here, the fermionic operator $O_F$ is antisymmetric, and satisfies the  constraint (reminiscent of the $\gamma_5$-hermiticity in lattice QCD)~\cite{Forini:2016sot,Bianchi:2016cyv}
 \begin{equation}\label{gamma5prop}
O_F^\dagger=\Gamma_5\,O_F\, \Gamma_5
\end{equation}
where $\Gamma_5$ is the following unitary, antihermitian matrix
\be\label{Gamma5}
\Gamma_5=\left(\begin{array}{cccc}
			0 			&  \mathbb{1}			&0& 0 \\
			-\mathbb{1}	&0					&0				&0\\
			0&    0					&0				& \mathbb{1}\\
			0			&0	&-\mathbb{1}		&0
          \end{array}\right)\,,
\qquad   \Gamma_5^\dagger \Gamma_5=\mathbb{1}  \qquad \Gamma_5^\dagger=-\Gamma_5\,,
\ee
which ensures that  $\det O_F$ is \emph{real} and \emph{non-negative}. %In particular, it can be seen that if $\lambda$ is an eigenvalue, then also  $-\lambda^*$ is an eigenvalue. Since OF is skew-symmetric also ?? and ?? 
While the absence of a complex phase allows us to eliminate a systematic error of our previous analysis~\cite{toappear},  it is not enough to make the Pfaffian definite positive, implying that the model may still suffer a sign problem.
One can check that -- in the case of generally complex eigenvalues $\lambda$ -- the antisymmetry and the $\Gamma_5$-hermiticity \eqref{gamma5prop} ensure a spectrum characterized by \emph{quartets} $(\lambda, -\lambda^*, -\lambda, \lambda^*)$.  
%, which would give a definite positive Pfaffian. 
One can then define the Pfaffian on the starting configuration as the product $(\lambda\,\lambda^*)$ for each quartet, which would provide sign flips in ${\rm Pf} O_F$. 
However, for purely imaginary or purely real eigenvalues, the disposition in quartets is no longer enforced by \eqref{gamma5prop}  and indeed may not happen, leaving a spectrum of pairs $(\lambda,-\lambda)$ with no degeneracy. A numerical study of the spectrum of $O_F$ appears to indicate that the disposition in quartets would occur if the $A$-terms in~\eqref{OF} -- defining Yukawa-like terms -- were vanishing, see Figure \ref{fig:spectrum_ferm} left, while for $A\neq 0$ (on the right) purely imaginary eigenvalues may appear, with no degeneracy. One should notice that such purely imaginary eigenvalues appear also when auxiliary fields are set to zero  - and thus the only non-vanishing $A$-term is  the one present in the original Lagrangian, before linearization -- suggesting that the sign ambiguity cannot be tamed by a suitably-enough choice of auxiliary fields.  
\begin{figure}[h]
   \centering
 \includegraphics[scale=0.6]{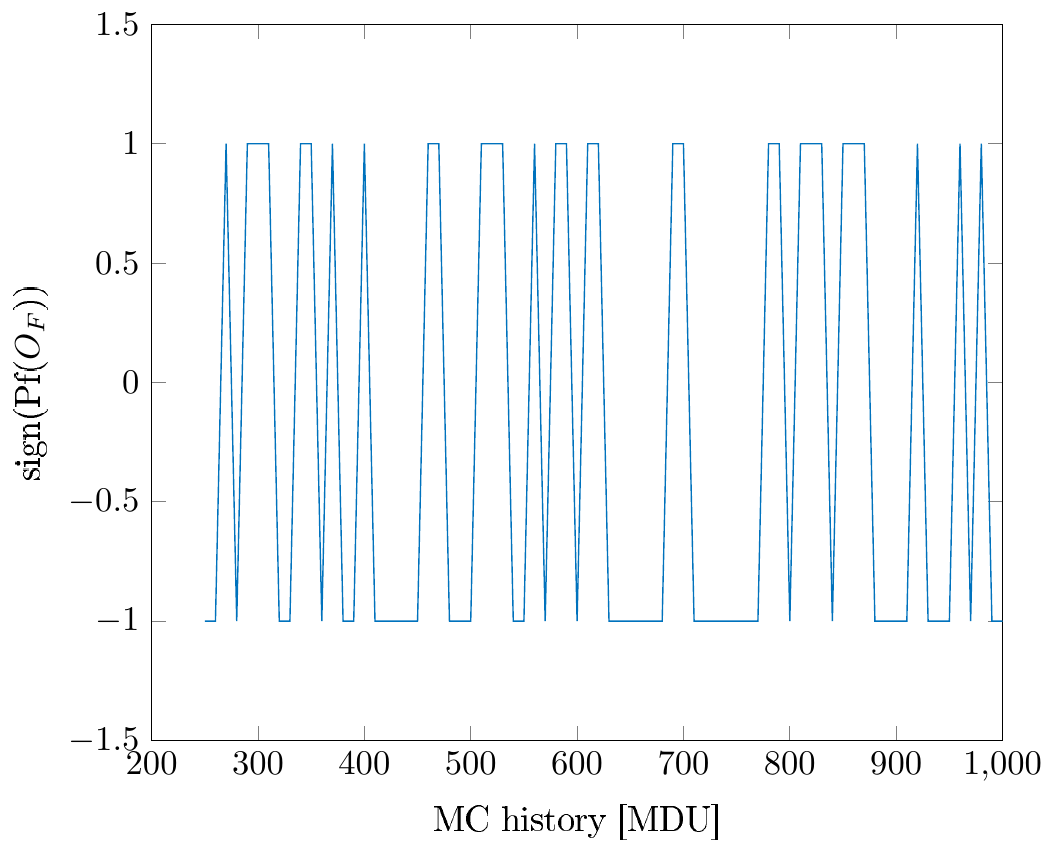}
  \includegraphics[scale=0.6]{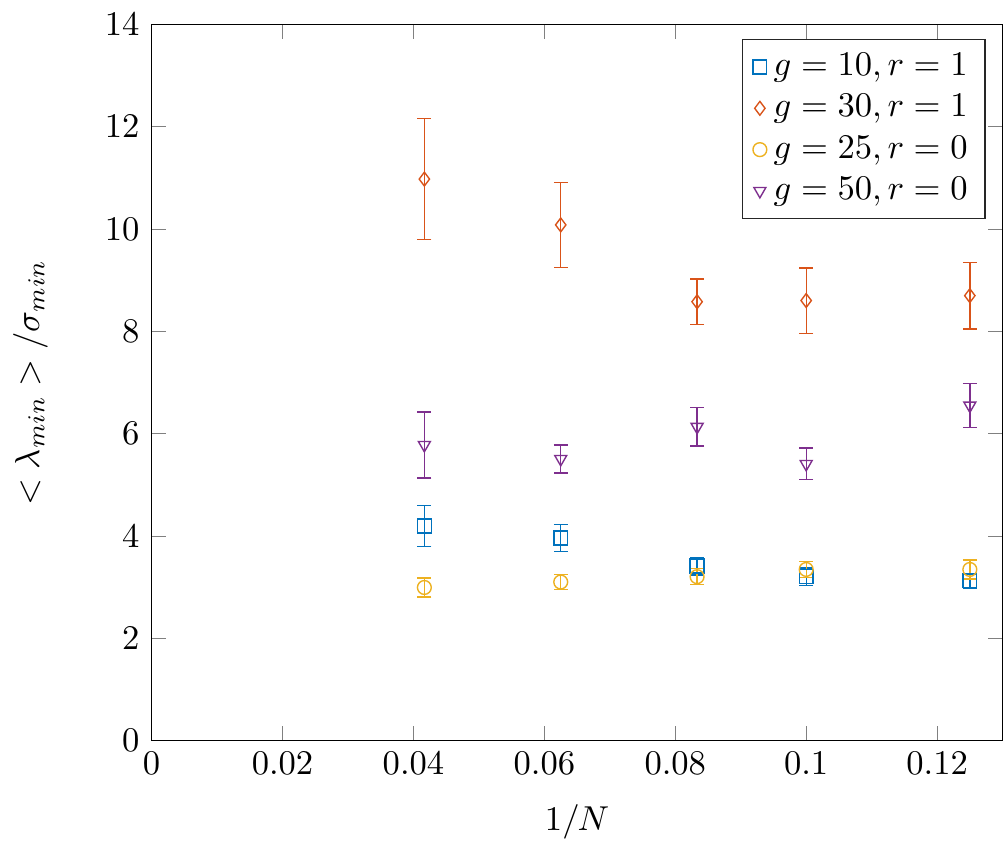}
 \caption{{\bf Left panel:} Study of the sign of ${\rm Pf}(O_F)$ %through $W={\rm Pf}(O_F)/(O_F^\dagger O_F+\mu^2)^{\frac{1}{4}}$,  
 for lattice size $L=8$ and $g=2$. 
 %The reweighting factor $W$ accounts for the introduction of a twisted mass term $\mu$ in $O_F$~\cite{Luscher:2008tw} %(via $O_F'=O_F+i\,\mu\,\Gamma_5$) 
 %so to shift the eigenvalues of $O_F$ away from zero ensuring a better convergence of the conjugate gradient solver~\cite{toappear}. 
 {\bf Right panel:}  The lowest  eigenvalue $\lambda_{\rm min}$ for the squared fermionic operator $O_F^\dagger O_F$  
  is well separated from zero, a statement which then also for $O_F$. The variance is defined by $\sigma_{\rm min}=\langle \lambda_{\rm min}^2\rangle-\langle \lambda_{\rm min}\rangle^2$. In the region of parameters explored, no zero eigenvalues for $\det O_F$ occur, which  implies no sign flips, and therefore absence of sign problem, for its \emph{real} Pfaffian.}
\label{fig:sign_lambdamin}
\end{figure}
Figure \ref{fig:sign_lambdamin}, left panel, shows that the sign problem becomes severe for values of the coupling $g\sim 2$. Interestingly, the presence (case $r=1$) of a Wilson-like discretization~\cite{Bianchi:2016cyv} appears to shift its occurrence to lower $g$~\cite{toappear}.  
One convenient way to answer the question of which region of the parameter space is free from a sign problem - and whether in such region information on the non-perturbative behavior of the system is obtainable -- is to study the lowest  eigenvalue  for the squared fermionic operator $O_F^\dagger O_F$  and identify a region of the parameter space where zero eigenvalues of such operator cannot occur (this ensures no zero eigenvalues for $O_F$).   This is done in the right panel of Figure \ref{fig:sign_lambdamin}, which shows that the smallest eigenvalues are clearly separated from zero for values of $g\geq 10$, where therefore sign flips cannot occur. It is interesting that this region safely includes $g=10$, at which (previous~\cite{Bianchi:2016cyv} and) current~\cite{toappear} simulations appear to detect the non-perturbative behavior, for example, of the (derivative of the) cusp anomaly as measured from the cusp action.
%\colb{sign flips in the Pfaffian (a continuous function of the eigenvalues) cannot occur at all. This is what is done in Fig. ??, where such distribution appears to be well separated from zero for the region  of couplings explored, $g\geq10$. In fact, it is reasonable to claim that this value  is  testing a  non-perturbative regime of the string sigma-model as $g=10$ measurements  do not belong to the perturbative realm of both observables. The latter is identifiable -- looking at the leading behavior and at the trend of perturbative corrections --  with $f?(g)/4\geq1$ and  $??$, while Fig. ?? shows .. and ??.}
% \begin{figure}[h]
%   \centering
% \includegraphics[scale=0.7]{nsigma-vs-N-1.pdf}
% \caption{The lowest part of the eigenvalue distribution of the quadratic (squared) fermion operator $O_F^\dagger O_F$ is well separated from zero. In the region of parameters explored, no zero eigenvalues for $\det O_F$ occur, which 
% implies no sign flips, and therefore absence of sign problem, for its \emph{real} Pfaffian. }
%\label{fig:lambda_min}
%\end{figure} 
 
\acknowledgments

\noindent We are grateful to S.~Catterall, R.~Roiban, D.~Schaich, K.~Skenderis, R.~Sommer and A.~Wipf  for very useful discussions.


\begin{thebibliography}{99}
 
 
    \bibitem{toappear}
L.~Bianchi, V.~Forini, B.~Leder, P.~T\"opfer, E.~Vescovi, to appear.
 
 
\bibitem{Roiban} 
  R.~W.~McKeown and R.~Roiban,
% ``The quantum $AdS_5 \times S^5$ superstring at finite coupling,''
  arXiv:1308.4875 [hep-th].
  %%CITATION = ARXIV:1308.4875;%%
  %1 citations counted in INSPIRE as of 07 Nov 2015

\bibitem{Forini:2016sot} 
  V.~Forini, L.~Bianchi, M.~S.~Bianchi, B.~Leder and E.~Vescovi,
  ``Lattice and string worldsheet in AdS/CFT: a numerical study,''
  PoS LATTICE {\bf 2015}, 244 (2016)
  [arXiv:1601.04670 [hep-lat]].
  %%CITATION = ARXIV:1601.04670;%%
  %4 citations counted in INSPIRE as of 27 Jan 2017

 
 \bibitem{Bianchi:2016cyv} 
  L.~Bianchi, M.~S.~Bianchi, V.~Forini, B.~Leder and E.~Vescovi,
  ``Green-Schwarz superstring on the lattice,''
  JHEP {\bf 1607}, 014 (2016)
  doi:10.1007/JHEP07(2016)014
  [arXiv:1605.01726 [hep-th]].
  %%CITATION = doi:10.1007/JHEP07(2016)014;%%
  %3 citations counted in INSPIRE as of 27 Jan 2017
%\cite{Forini:2016sot}


\bibitem{Catterall_physrept} 
  S.~Catterall, D.~B.~Kaplan and M.~Unsal,
  ``Exact lattice supersymmetry,''
  Phys.\ Rept.\  {\bf 484}, 71 (2009)
%  [arXiv:0903.4881 [hep-lat]].
  %%CITATION = ARXIV:0903.4881;%%
  %105 citations counted in INSPIRE as of 10 Nov 2015

  \bibitem{Schaich:2015ppr}
D.~Schaich, ``{Aspects of lattice $\mathcal N = 4$ supersymmetric Yang--Mills}'',
 %\href{http://pos.sissa.it/archive/conferences/251/242/LATTICE 2015_242.pdf}{\bf LATTICE 2015 (2015) 242}, 
PoS LATTICE {\bf 2015} 242 (2015)
arXiv:1512.01137 [hep-lat]

\bibitem{Bergner:2016sbv} 
  G.~Bergner and S.~Catterall,
``Supersymmetry on the lattice,''
  Int.\ J.\ Mod.\ Phys.\ A {\bf 31}, no. 22, 1643005 (2016)
  %doi:10.1142/S0217751X16430053
  [arXiv:1603.04478 [hep-lat]].
  %%CITATION = doi:10.1142/S0217751X16430053;%%
  %6 citations counted in INSPIRE as of 27 Jan 2017
%\cite{Catterall:2015zua}


%\cite{Schaich:2016jus}
\bibitem{Schaich:2016jus} 
  D.~Schaich, S.~Catterall, P.~H.~Damgaard and J.~Giedt,
  ``Latest results from lattice N=4 supersymmetric Yang--Mills,''
  PoS LATTICE {\bf 2016}, 221 (2016)
  [arXiv:1611.06561 [hep-lat]].
  %%CITATION = ARXIV:1611.06561;%%
  %4 citations counted in INSPIRE as of 27 Jan 2017
%\cite{Catterall:2016dzf}


%\cite{Berkowitz:2016jlq}
\bibitem{Berkowitz:2016jlq} 
  E.~Berkowitz, E.~Rinaldi, M.~Hanada, G.~Ishiki, S.~Shimasaki and P.~Vranas,
  %``Precision lattice test of the gauge/gravity duality at large-$N$,''
  Phys.\ Rev.\ D {\bf 94}, no. 9, 094501 (2016)
  doi:10.1103/PhysRevD.94.094501
  [arXiv:1606.04951 [hep-lat]].
  %%CITATION = doi:10.1103/PhysRevD.94.094501;%%
  %6 citations counted in INSPIRE as of 29 Jan 2017

  %\cite{Metsaev:2000yf}
\bibitem{MTMTT} 
  R.~R.~Metsaev and A.~A.~Tseytlin,
  ``Superstring action in $AdS_5\times S^5$. Kappa symmetry light cone gauge,''
  Phys.\ Rev.\ D {\bf 63}, 046002 (2001)
 % [hep-th/0007036]
 ; R.~R.~Metsaev, C.~B.~Thorn and A.~A.~Tseytlin,
  %``Light cone superstring in AdS space-time,''
  Nucl.\ Phys.\ B {\bf 596}, 151 (2001)
 % [hep-th/0009171].


   %\cite{Beisert:2010jr}
%\bibitem{Beisert_review} 
%  N.~Beisert {\it et al.},
%  ``Review of AdS/CFT Integrability: An Overview,''
%  Lett.\ Math.\ Phys.\  {\bf 99}, 3 (2012)
  %[arXiv:1012.3982 [hep-th]].
  %%CITATION = ARXIV:1012.3982;%%
  %519 citations counted in INSPIRE as of 10 Nov 2015

%\cite{Pestun:2007rz}
%\bibitem{Pestun:2007rz} 
%  V.~Pestun,
  %``Localization of gauge theory on a four-sphere and supersymmetric Wilson loops,''
%  Commun.\ Math.\ Phys.\  {\bf 313}, 71 (2012)
%  doi:10.1007/s00220-012-1485-0
%  [arXiv:0712.2824 [hep-th]].
  %%CITATION = doi:10.1007/s00220-012-1485-0;%%
  %580 citations counted in INSPIRE as of 21 Dec 2015
  


%\cite{Maldacena:1998im}
%\bibitem{MaldaWL} 
%  J.~M.~Maldacena,
%  ``Wilson loops in large N field theories,''
%  Phys.\ Rev.\ Lett.\  {\bf 80}, 4859 (1998)
%  [hep-th/9803002].
  %%CITATION = HEP-TH/9803002;%%
  %1278 citations counted in INSPIRE as of 10 Nov 2015

%\cite{Korchemsky:1992xv}
%\bibitem{Korchemsky} 
%  G.~P.~Korchemsky and G.~Marchesini,
%  ``Structure function for large x and renormalization of Wilson loop,''
%  Nucl.\ Phys.\ B {\bf 406}, 225 (1993)
%  [hep-ph/9210281].
  %%CITATION = HEP-PH/9210281;%%
  %264 citations counted in INSPIRE as of 10 Nov 2015


\bibitem{Giombi} 
%\cite{Giombi:2009gd}
  S.~Giombi, R.~Ricci, R.~Roiban, A.~A.~Tseytlin and C.~Vergu,
  ``Quantum AdS(5) x S5 superstring in the AdS light-cone gauge,''
  JHEP {\bf 1003}, 003 (2010)
 % [arXiv:0912.5105 [hep-th]].
  %%CITATION = ARXIV:0912.5105;%%
  %23 citations counted in INSPIRE as of 10 Nov 2015





%\bibitem{Giombi:2010bj} 
%  S.~Giombi, R.~Ricci, R.~Roiban and A.~A.~Tseytlin,
%  ``Quantum dispersion relations for excitations of long folded spinning superstring in $AdS_5\times S^5$,''
%  JHEP {\bf 1101}, 128 (2011)
%  doi:10.1007/JHEP01(2011)128
%  [arXiv:1011.2755 [hep-th]].
  %%CITATION = doi:10.1007/JHEP01(2011)128;%%
  %10 citations counted in INSPIRE as of 22 Dec 2015
 
 %\cite{Wolff:2003sm}
%\bibitem{Wolff:2003sm} 
%  U.~Wolff [ALPHA Collaboration],
  %``Monte Carlo errors with less errors,''
 % Comput.\ Phys.\ Commun.\  {\bf 156}, 143 (2004)
 % [Comput.\ Phys.\ Commun.\  {\bf 176}, 383 (2007)]
  %doi:10.1016/S0010-4655(03)00467-3, 10.1016/j.cpc.2006.12.001
  %[hep-lat/0306017].
  %%CITATION = doi:10.1016/S0010-4655(03)00467-3, 10.1016/j.cpc.2006.12.001;%%
  %135 citations counted in INSPIRE as of 15 Jan 2016
 
 
%\bibitem{AldayMaldacena}
%L.~F.~Alday and J.~M.~Maldacena, %``Comments on operators with large spin,'' 
%JHEP 0711, 019 (2007) [arXiv:0708.0672 [hep-th]]; Z.~Bajnok, J.~Balog, B.~Basso, G.~P.~Korchemsky and L.~Palla, %ñScaling function in AdS/CFT from the O(6) sigma model,î 
%Nucl. Phys. B 811, 438 (2009) 
%[arXiv:0809.4952 [hep-th]].


%\cite{Beisert:2006ez}
\bibitem{BES} 
  N.~Beisert, B.~Eden and M.~Staudacher,
  ``Transcendentality and Crossing,''
  J.\ Stat.\ Mech.\  {\bf 0701}, P01021 (2007)
 % [hep-th/0610251].
  %%CITATION = HEP-TH/0610251;%%
  %598 citations counted in INSPIRE as of 07 Nov 2015

  \bibitem{Basso} 
  B.~Basso,
  ``Exciting the GKP string at any coupling,''
  Nucl.\ Phys.\ B {\bf 857}, 254 (2012)
  [arXiv:1010.5237 [hep-th]].
  %%CITATION = ARXIV:1010.5237;%%
  %49 citations counted in INSPIRE as of 10 Nov 2015
  
  %\cite{Catterall:2014vga}




\bibitem{Catterall:2015zua} 
  S.~Catterall,
``Fermion mass without symmetry breaking,''
  JHEP {\bf 1601}, 121 (2016)
  doi:10.1007/JHEP01(2016)121
  [arXiv:1510.04153 [hep-lat]].
  %%CITATION = doi:10.1007/JHEP01(2016)121;%%
  %7 citations counted in INSPIRE as of 27 Jan 2017
 
   %\cite{Luscher:2008tw}
%\bibitem{Luscher:2008tw} 
%  M.~Luscher and F.~Palombi,
  %``Fluctuations and reweighting of the quark determinant on large lattices,''
 % PoS LATTICE {\bf 2008}, 049 (2008)
  %[arXiv:0810.0946 [hep-lat]].
  %%CITATION = ARXIV:0810.0946;%%
  %24 citations counted in INSPIRE as of 25 Jan 2017

\bibitem{Catterall:2016dzf} 
  S.~Catterall and D.~Schaich,
  ``Novel phases in strongly coupled four-fermion theories,''
  arXiv:1609.08541 [hep-lat].
  %%CITATION = ARXIV:1609.08541;%%
%\cite{Bergner:2016sbv}

\end{thebibliography}
\end{document}